\documentclass[twocolumn,prl,showpacs,superscriptaddress,amssymb,amsmath]{revtex4}
\usepackage{lmodern}
\usepackage[T1]{fontenc}
\usepackage{graphicx}
\usepackage{epstopdf}
\usepackage{bm}
\usepackage{hyperref}
\usepackage{wasysym}

\hypersetup{%
  pdfpagemode=UseNone, 
  pdfstartpage=1,
  pdfstartview=FitH,
  pdfmenubar=true,
  pdftoolbar=true,
  colorlinks = true,
  linkcolor=blue,
  citecolor=blue,
  bookmarksopen=false
}

\begin{document}

\title{Controlling magnetic Feshbach resonances \\in polar open-shell
  molecules with non-resonant light} 

\author{Micha\l~Tomza}
\affiliation{Faculty of Chemistry, University of Warsaw, 
  Pasteura 1, 02-093 Warsaw, Poland}
\affiliation{Instituto 'Carlos I' de F\'isica
    Te\'orica y Computacional and Departamento de F\'isica At\'omica,
    Molecular y Nuclear, Universidad de Granada, 18071 Granada, Spain}
\affiliation{Theoretische Physik, Universit\"at Kassel, 
  Heinrich-Plett-Str. 40, 34132 Kassel, Germany}
\author{Rosario  Gonz\'alez-F\'erez}
\affiliation{Instituto 'Carlos I' de F\'isica
    Te\'orica y Computacional and Departamento de F\'isica At\'omica,
    Molecular y Nuclear, Universidad de Granada, 18071 Granada, Spain}
\affiliation{The Hamburg Center for Ultrafast Imaging, University of
  Hamburg, 22761 Hamburg, Germany}
\author{Christiane P. Koch}
\email{christiane.koch@uni-kassel.de}
\affiliation{Theoretische Physik, Universit\"at Kassel, 
  Heinrich-Plett-Str. 40, 34132 Kassel, Germany}
\author{Robert Moszynski}
\affiliation{Faculty of Chemistry, University of Warsaw, 
  Pasteura 1, 02-093 Warsaw, Poland}

\date{\today}

\begin{abstract}
Magnetically tunable Feshbach resonances for polar paramagnetic
ground-state diatomics
are too narrow to allow for magnetoassociation starting
from trapped, ultracold atoms. We show that non-resonant light can 
be used to engineer the Feshbach resonances in their position and
width. For non-resonant field intensities of the order of
$10^9\,$W/cm$^2$, we find the width to be increased by three 
orders of magnitude, reaching a few Gauss. This opens the way for
producing ultracold molecules with sizeable electric and magnetic
dipole moments and thus for many-body quantum simulations with such
particles. 
\end{abstract}

\maketitle

Ultracold polar molecules are predicted to probe fundamental
physics~\cite{DeMillePRL08} and realize a wealth of many-body
phenomena such as exotic quantum phases~\cite{YaoPRL13}.
They are thus attracting significant interest in both the AMO and
condensed matter communities~\cite{CarrNJP09}. 
Polar alkali dimers have already been produced in their
absolute internal ground state close to quantum
degeneracy~\cite{NiScience08}, opening the way toward ultracold
chemistry~\cite{OspelkausSci10,MirandaNatPhys11} and quantum
simulation~\cite{MicheliNatPhys06}. Contrary to ground-state alkali
dimers which are  closed-shell, diatomics 
consisting of an open-shell and a closed-shell atom possess an 
unpaired electron, endowing the molecule with spin structure and 
a significant magnetic dipole moment. 
Since these molecules have both electric and magnetic dipoles, 
they are supreme candidates for 
creating topologically ordered states~\cite{BlochRMP08},
investigating collective spin excitations~\cite{KremsNJP10}
and realizing lattice-spin models~\cite{MicheliNatPhys06}. 
While numerous ultracold mixtures of open-shell alkali and
closed-shell Yb or Sr atoms have already been
studied
experimentally~\cite{NemitzPRA09,BaumerPRA11,DoylePRL11,IvanovPRL11,Schreck13,AokiPRA13}, 
magnetoassociating the atoms into molecules has remained an
elusive goal. 

The most successful and widely used routes to producing ultracold 
dimers utilize magnetically tunable Feshbach
resonances (FRs)~\cite{JulienneRMP06b,JulienneRMP10} where the hyperfine
interaction couples a scattering state to a bound molecular level. 
Somewhat unexpectedly, FRs have been predicted 
for diatomics with a $^2\Sigma$ ground electronic state
such as RbSr and LiYb~\cite{ZuchowskiPRL10,BruePRL12}. 
The resonances are caused by a modification of the alkali atom's
hyperfine structure due to the presence of the other
atom~\cite{ZuchowskiPRL10} or by creating a hyperfine coupling due to
the alkali atom polarizing the nuclear spin
density of fermionic Yb~\cite{BruePRL12}. 
However, the width of these resonances 
does not exceed a few milli-Gauss. This renders their use in
magnetoassociation very difficult, if not impossible. A different kind
of FR for a closed-shell/open-shell mixture has
recently been observed, with one of the atoms in an electronically
excited state~\cite{KatoPRL13,KhramovPRL14}. In this case, the FR is
induced by the anisotropy of the interaction between $S$-state
and $P$-state atoms. 
Due to the finite excited state lifetime, such a FR
is not ideally suited for making molecules. It suggests, however, to
harness an anisotropic interaction for magnetoassociation. 

Here we show that non-resonant light, which universally couples to the
polarizability anisotropy of a molecule, induces
FRs and modifies their position and width. This is due
to the non-resonant light changing the background scattering length and 
altering the differential magnetic susceptibility.  
Our approach is related to dc electric field control of polar
molecules~\cite{KremsPRL06,LiPRA07,MarcelisPRL08} but comes with much
more favorable requirements in terms of experimental feasibility. 
We find widths of a few Gauss for non-resonant field intensities of
the order $10^9\,$W/cm$^2$ for 
a wide range of polar open-shell molecular species. 
Widths of a few Gauss are sufficient for magnetoassociation.
Non-resonant field control thus paves the way to producing ultracold 
particles with sizeable electric and magnetic dipole moment. 

Magnetoassociation 
can employ an adiabatic ramp of the magnetic field
across the resonance or a time-dependent magnetic or radio-frequency
(rf) field that drives a transition from a scattering state to a molecular
level~\cite{JulienneRMP10}. 
These two routes imply different requirements on the characteristics
of the resonance. In both cases, a broad FR is needed.
Adiabatic passage additionally
requires a large product of width, $\Delta$, and background scattering 
length, $a_{bg}$. This is seen in the 
atom-molecule conversion efficiency, given by the Landau-Zener
formula $1-\exp\left[-\eta n\frac{\hbar}{\mu}
\left|\frac{a_{bg}\Delta}{\dot{B}}\right|\right]$ 
with $n$ the atomic number density, $\dot{B}$ the magnetic field ramp
speed,  $\mu$ the reduced mass and
$\eta$ a dimensionless prefactor~\cite{HodbyPRL05}.
Using Fermi's Golden Rule, the resonance width $\Delta$ can be
estimated, 
\begin{equation}\label{eq:Delta}
\Delta \sim \frac{|\langle v|H|k \rangle|^2}{a_{{bg}}\delta\chi}\,,
\end{equation}
in terms of the coupling $\langle v|H|k \rangle$
between molecular level $|v\rangle$ and
scattering states $|k\rangle$, the background scattering
length $a_{bg}$, and the differential magnetic susceptibility,
$\delta\chi$~\cite{BruePRA13}.
The latter is simply the difference in slope 
of the bound and continuum energies as function of magnetic field at
resonance. 
When the background scattering length $a_{bg}$ is larger than the mean
scattering length $\bar{a}$ ($\bar a\approx0.48(2\mu
C_6/\hbar)^{1/4}$ with $C_6$ the dispersion coefficient), the 
coupling $|\langle v|H|k \rangle|$ becomes proportional to
$a_{{bg}}$. The width is then determined by background scattering
length and differential magnetic susceptibility, 
$\Delta \sim {a_{{bg}}}/{\delta\chi}$~\cite{BruePRA13}.
The key point of our proposal is that both $\delta\chi$ and $a_{bg}$
can be tuned by applying a non-resonant field. This leads to
significant changes in the resonance width $\Delta$ and the
adiabaticity parameter $|a_{bg}\Delta|$.

The Hamiltonian describing the relative nuclear motion of 
an open-shell $^2S$
atom, $a$, and a closed-shell $^1S$ atom, $b$, reads
\begin{equation}\label{eq:Ham}
  \hat{H}=\frac{\hbar^2}{2\mu}\left(-\frac{1}{r}\frac{d^2}{dr^2}r
    +\frac{\hat{L}^2}{r^2}\right)+\hat H_{a}+\hat H_{b}+V(r,\theta)\,,
\end{equation}
where $r$ is the interatomic separation,  
$\hat{L}$ the rotational angular momentum operator, and
$\theta$ the angle between the molecular axis and 
the space-fixed $Z$-axis. The atomic Hamiltonian including
Zeeman and hyperfine interactions is given by 
\begin{equation}
  \label{eq:Hatom}
  \hat{H}_{j}=\zeta_{j}\hat{i}_{j}\cdot\hat{s}_{j}
  +\left(g_e\mu_{{B}}\hat{s}_{j,z}+g_{j}\mu_{{N}}\hat{i}_{j,z}\right)B\,,
\end{equation}
with $\hat{s}_{j}$ and $\hat{i}_{j}$ the electron and
nuclear spin operators, $g_{e/j}$  the electron and
nuclear $g$ factors, and $\mu_{B/N}$ the Bohr and nuclear
magnetons.  $\zeta_{j}$ denotes the hyperfine coupling constant.
For a fermionic closed-shell $^1S$ atom, Eq.~(\ref{eq:Hatom})
reduces to the nuclear Zeeman term, 
whereas for a bosonic one it is zero. 
The interatomic interaction reads
\begin{eqnarray}
  \label{eq:V} \nonumber
  V(r,\theta)&=&V_{X^2\Sigma^+}(r)+\Delta\zeta_{a}(r)\hat{i}_{a}\cdot \hat{s}_{a}\\
  &&-\frac{I}{2\epsilon_0c}\left(\alpha_{\perp}(r)+\Delta\alpha(r)\cos^2\theta\right) 
\end{eqnarray}
for  magnetic and non-resonant laser fields parallel to the
space-fixed $Z$-axis.
$V_{X^2\Sigma^+}(r)$ is the potential energy curve for the
$X^2\Sigma^+$ ground electronic state, and $\Delta\zeta_{a}(r)$ the
interaction-induced variation 
of the hyperfine coupling~\cite{ZuchowskiPRL10,BruePRL12}. 
The molecular static 
polarizability with perpendicular component $\alpha_{\perp}(r)$
and anisotropy $\Delta\alpha(r)$ couples to 
non-resonant light of intensity $I$, linearly polarized along the
space-fixed $Z$-axis. We omit 
spin-rotation couplings as well as the
coupling resulting from a non-zero nuclear spin of a fermionic
closed-shell atom since they are significantly smaller than
$\Delta\zeta_a(r)$. 

We focus on RbYb  for which spectroscopic and
\textit{ab initio} data for the interaction potential are
available~\cite{BorkowskiPRA13}. 
The $r$-dependent isotropic and anisotropic polarizabilities are 
calculated using state of the art coupled cluster methods, small-core
energy consistent 
pseudopotentials, and large basis sets~\cite{TomzaFRlong}. They
perfectly agree with Silberstein's formula~\cite{HeijmenMP96,JensenJCP02}
evaluated for the atomic 
polarizabilities of Ref.~\cite{DerevienkoADNDT10}.
The interaction-induced variation of the hyperfine coupling, 
$\Delta\zeta_{a}(r)$, is taken from Ref.~\cite{BruePRA13}. 
The total scattering wave function is constructed in an uncoupled
basis set,
$|i_{a},m_{i,a}\rangle|s_{a},m_{s,a}\rangle|L,m_{L}\rangle$ with $m_j$
the projection of angular momentum $j$ on the space-fixed $Z$ axis,
assuming the projection of the total angular momentum of rubidium
$m_{f}=m_{i,a}+m_{s,a}$ to be conserved.
The coupled channels equations are solved using a renormalized Numerov 
propagator~\cite{JohnsonJCP78}. 
The scattering lengths and elastic cross sections are obtained from
the $S$ matrix for the entrance channel, 
$a=(1-S_{11})/(1+S_{11})/(ik)$ and
$\sigma_{el}=\pi|1-S_{11}|/k^2$, 
with $k=\sqrt{2\mu E}/\hbar$ and $E$ the collision energy, assumed to
be 100$\,$nK. The resonance width $\Delta$ is determined by fitting the
scattering length to
$a(B)=a_{bg}(1-\Delta/(B-B_{res}))$ \cite{JulienneRMP06b,JulienneRMP10}. 

\begin{figure}[tb]
  \begin{center}
    \includegraphics[width=\columnwidth]{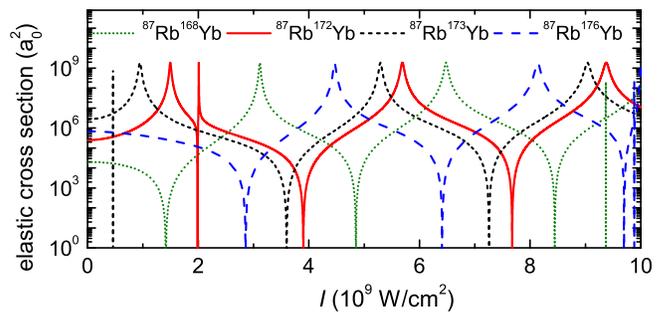}
  \end{center}
  \caption{(Color online) Non-resonant light control of scattering
    properties: Elastic cross section as a function of
    the non-resonant light intensity ($E/k_B=100\,$nK, $B=0$).  
}
\label{fig:cross}
\end{figure}
\begin{figure}[tb]
\begin{center}
  \includegraphics[width=\columnwidth]{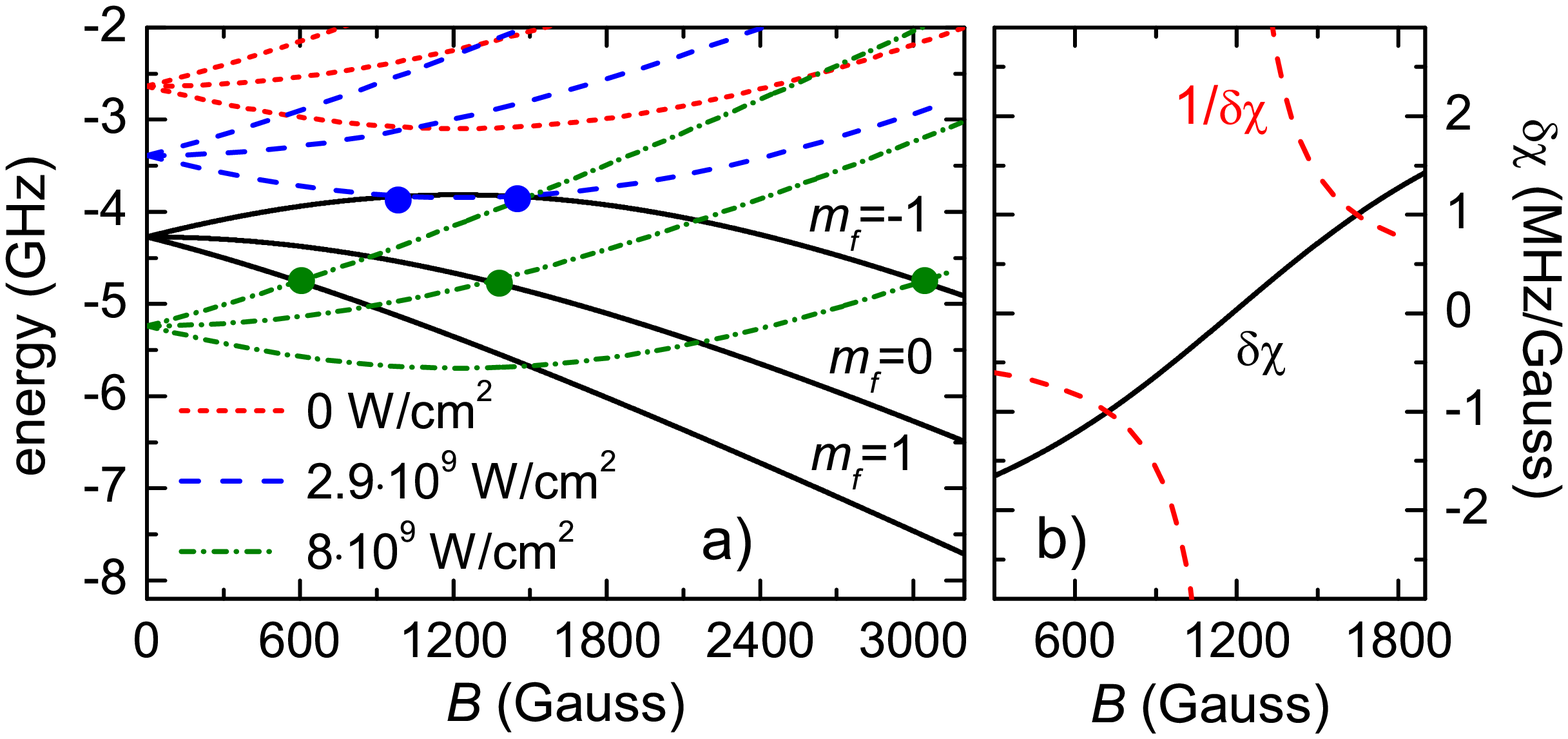}
\end{center}
\caption{(Color online) Creating new FR with non-resonant light: 
  a) Atomic thresholds (solid black  lines) start to cross 
  molecular levels (dashed lines) as the non-resonant light
  shifts the level positions ($^{87}$Rb$^{176}$Yb with $|m_f|\le
  i_a-1/2$). 
  The dots  indicate the position of the new FR. b) The level
  shifts are accompanied by a 
  variation of the differential magnetic susceptibility $\delta\chi$ vs
  magnetic field ($m_f=-1$, $I=0$). 
}
\label{fig:levels}
\end{figure}

Non-resonant light modifies the energies of rovibrational
levels and scattering states
alike~\cite{LemeshkoPRL09,LemeshkoJPCA10,GonzalezPRA12,TomzaMP13}. The 
latter implies control of scattering properties
such as the cross sections. This is illustrated by 
Fig.~\ref{fig:cross} which displays a series of maxima and minima of the
elastic cross section as a function of non-resonant field intensity. The
maxima correspond to a large absolute value of $a_{bg}$ and occur when
a scattering state becomes bound; the minima indicate
non-interacting atoms. Broad maxima of the elastic cross section are
observed when an $s$-wave 
scattering state is pushed below threshold, whereas the narrow features
in Fig.~\ref{fig:cross} are caused by higher partial waves.
New FRs are created by the non-resonant light shifting bound
levels. This happens when a bound level crosses the atomic threshold of
a different hyperfine level as indicated by the
dots in Fig.~\ref{fig:levels}a). 
New resonances, higher than
$s$-wave, may also be induced by mixing partial waves or by 
spin-rotation coupling between higher partial waves. 
The non-resonant field dependence of the background scattering length
observed in Fig.~\ref{fig:cross} and the creation of new FR due to the
non-resonant light shown in
Fig.~\ref{fig:levels} together with Eq.~\eqref{eq:Delta} suggest
three mechanisms to increase the width of FRs: (i)
$\delta\chi \to 0$, (ii) $|a_{bg}|\to \infty$, and (iii) $|a_{bg}|\to
0$. In case~(i), $|a_{bg}\Delta|$ becomes
large unless it coincides with case (iii), and large $|a_{bg}\Delta|$
is guaranteed in case~(ii). Then both  
adiabatic ramping across the resonance and rf association are possible. In
contrast, $|a_{bg}\Delta|$ will always stay small in case~(iii),
preventing adiabatic passage. Since adiabatic ramping is the most
popular technique for magnetoassociation, we focus on cases~(i) and
(ii) here and will report on case (iii) elsewhere~\cite{TomzaFRlong}. 

\begin{figure}[tb]
\begin{center}
  \includegraphics[width=\columnwidth]{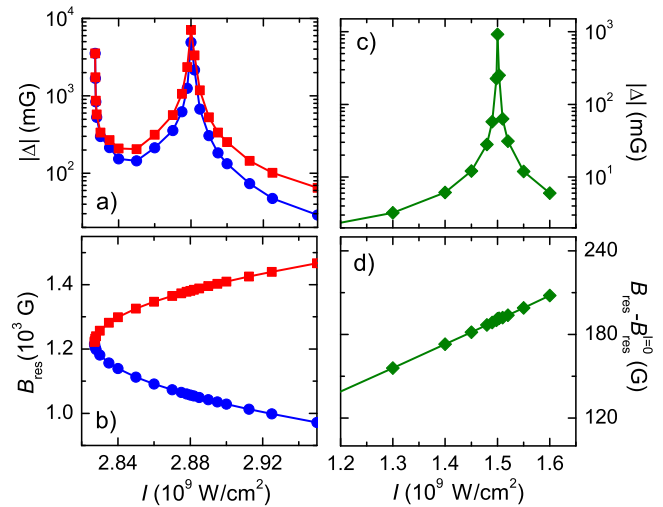}
\end{center}
\caption{(Color online)  
  Controlling the width of a FR by modifying $\delta\chi$ (a,b):
  Resonance width $\Delta$ and 
  resonance position $B_{res}$ vs non-resonant light intensity
  for $^{87}$Rb$^{176}$Yb and the pair of resonances 
  indicated by blue dots in Fig.~\ref{fig:levels} ($m_f=-1$, 
  $B_{res}=1219\,$G).  
  Controlling the width of a FR by tuning $a_{bg}$ to large values
  (c,d):  Resonance width $\Delta$ and change in 
  resonance  position $B_{res}-B^{I=0}_{res}$ vs non-resonant
  light intensity for $^{87}$Rb$^{172}$Yb ($m_f=1$, $B_{res}^{I=0}=1592\,$G). 
} 
\label{fig:control}
\end{figure}
We find that case (i) yields the largest widths. This is illustrated
in  Fig.~\ref{fig:control}a) for $^{87}$Rb$^{176}$Yb: A pair of
resonances is created when the molecular level crosses an atomic
threshold close  to the maximum of its
magnetic field dependence, cf. blue dots in
Fig.~\ref{fig:levels}a). The resonances come with a 
very large width $\Delta$, of the order of a few Gauss, cf. the left peak in
Fig.~\ref{fig:control}a), and are separated by several Gauss (by 6$\,$G for
example for $\Delta\approx 3\,$G). 
The large width is rationalized by the broad pole of
$1/\delta\chi$ shown in Fig.~\ref{fig:levels}b) which enters
Eq.~\eqref{eq:Delta}. Not only the width $\Delta$ but 
also the adiabaticity parameter
 $|a_{bg}\Delta|$ is found to be large, of the order of
10$\,$a$_0\cdot$G, whereas the background scattering length remains
comparatively small, of the order of 10$\,$a$_0$.
A second peak of the width $\Delta$, of the order of 10$\,$G,
 is observed in Fig.~\ref{fig:control}a), at
$I=2.88\cdot10^9\,$W/cm$^2$. It is caused by $a_{bg}$ 
going to zero, which can be inferred from the corresponding minimum of
the blue dashed curve in Fig.~\ref{fig:cross}.
The joint occurrence of $\delta\chi\to 0$ and $a_{bg}\to
0$ is a coincidence. 
As can be seen in  Fig.~\ref{fig:control}a) and
b), such a coincidence leads to particularly broad resonances
for a range of non-resonant field
intensities which at the same time are separated by several hundreds
Gauss. However, due to $a_{bg}\to 0$, the adiabaticity parameter
$|a_{bg}\Delta|$ remains small. 
The adiabaticity parameter is guaranteed to be large in 
case (ii) when the non-resonant field is used to tune the background
scattering length to very large values. This is illustrated in 
Fig.~\ref{fig:control}c) and d). 
The maximum width $\Delta$ which is not limited in theory
will depend on the stability of the non-resonant field intensity 
in practice. For 
example, an increase by $10^3$ requires intensity stabiliziation of
the order $10^{-3}$ to $10^{-4}$. 
The actual value of $\Delta$ that can be obtained also depends 
on the field-free width. But even for very narrow resonances,
with the field-free $\Delta$ below 1$\,$mG, the engineered width easily reaches 
100$\,$mG, as demonstrated by 
Fig.~\ref{fig:control}c). 

We find non-resonant light intensities of the order of
$10^9\,$W/cm$^2$ to be sufficient to 
create FRs for all isotopologues of RbYb. The smallest 
intensity is required for diatomics with a molecular level
just above the atomic threshold since the  non-resonant field always
lowers the energy in the electronic ground state~\cite{TomzaMP13}.  
For example, a pair of broad resonances as shown in 
Fig.~\ref{fig:control}a,b) is also observed for $^{85}$Rb${}^{170}$Yb 
(with $\Delta>0.5\,$G at $I=1.29\cdot10^9\,$W/cm$^2$). 
When only the rubidium isotope is exchanged, the dependence on the
non-resonant light intensity remains essentially unchanged compared to
Fig.~\ref{fig:control}a,b). Of course, different hyperfine levels may
come into play, e.g., $m_f=-2$ or $m_f=-1$, which imply 
different magnetic fields ($B_{res}=722\,$G and  $B_{res}=361\,$G,
respectively, for $^{85}$Rb$^{176}$Yb). The left peak of $\Delta$ in
Fig.~\ref{fig:control}a) and the associated increase in
$|a_{bg}\Delta|$ is found for all RbYb isotopologues. The right
peak corresponds to a coincidence of case~(i) 
with case~(iii) 
and is specific to
$^{87}$Rb$^{176}$Yb. Case~(i) may coincide also with case~(ii). This happens
for $^{87}$Rb$^{174}$Yb, yielding an adiabaticity parameter
$|a_{bg}\Delta|$ of the order of 100$\,$a$_0\cdot$G. Case
(ii), i.e., large $a_{bg}$, is most easily realized for molecules
with a large and negative field-free background scattering length $a_s$. 
For $^{87}$Rb${}^{172}$Yb shown in Fig.~\ref{fig:control}c,d) for
example $a_s=-131\,a_0$~\cite{MunchowPhD12}. Another good candidate
for case (ii) is $^{87}$Rb${}^{173}$Yb (with
$a_s=-431\,a_0$~\cite{MunchowPhD12}). 

The three mechanisms are generally applicable due to the universal
coupling to non-resonant light. Notably, we find    
the characteristics of controlling the resonance width by tuning the 
background scattering length as shown in Fig.~\ref{fig:control}c,d)
to be common to all $^2\Sigma$ molecules.
When considering closed-shell/open-shell mixtures other than RbYb,
different strengths of both magnetic field and non-resonant light
might, however, be required. For example, LiYb has a smaller reduced
mass than RbYb and Li a smaller polarizability than Rb which implies a
larger non-resonant field intensity.
The magnetic field strength for which a molecular level crosses the
atomic threshold close to the maximum of its magnetic field
dependence, relevant for case (i), is determined by the hyperfine
splitting~\cite{BruePRA13}.  It is thus smaller
for mixtures involving Li, Na or K and larger for those involving Cs 
instead of Rb. Prospects are best for RbSr and 
CsYb~\cite{TomzaFRlong} for which the interaction induced variation of
the hyperfine structure and the polarizabilities are largest. 
Together with the tunability of the field-free background scattering
length by choice of the Yb isotope, this makes CsYb in particular another very 
promising candidate. 

When tuning non-resonant light and magnetic field for interspecies
magnetoassociation, undesired losses may occur due to accidentally 
hitting an intraspecies FR or shape resonance. For example, for RbYb,
depending on the isotope, one to three shape resonances are observed
for Yb$_2$ at  non-resonant field intensities of the order of
$10^{9}\,$W/cm$^2$. The shape 
resonance found for $^{176}$Yb$_2$ at $I=3.05\cdot 10^9\,$W/cm$^2$ is
sufficiently far from  $I=2.83\cdot 10^9\,$W/cm$^2$, for which 
the width of the $^{87}$Rb$^{176}$Yb FR is increased 
to several Gauss, cf. Fig.~\ref{fig:control}a). 
The separation is even slightly larger for $^{174}$Yb$_2$. 
Alkali intraspecies FRs are found to be shifted in
position by the non-resonant field. If, as the result, an 
intraspecies FR is moved too close to the interspecies
one, a different Yb isotope should be selected. Similarly,
selection of the hyperfine level provides a solution, if a shape
resonance approaches the interspecies FR too closely,
for example for CsYb. Perturbations due to intraspecies resonances can
thus be avoided. Such losses do not occur altogether when working in a
double-species Mott insulator 
state~\cite{HaraJPSJ14}. 

Our proposal for non-resonant light controlled magnetoassociation
requires intensities of the order of $10^9\,$W/cm$^2$ and magnetic
fields of the order of 1000$\,$G. These requirements are
within current experimental capabilities. Intensities of the order
10$^9\,$W/cm$^2$ can be  achieved using intracavity beams with spot
sizes of about $10\,\mu$m and powers of the order of $1\,$kW. Such 
spot sizes could be desirable for creating an additional trap. 
Larger spot sizes, up to  $100\,\mu$m, are possible when using an optical buildup
cavity~\cite{YostOE11,CingozNat12}. The required intensity can
be stabilized at a level of 0.001, but even $10^{-4}$ should
be reachable with refined feedback techniques. 
Magnetic fields can be stabilized at the level
$10^{-5}$-$10^{-6}$~\cite{ZurnPRL13} 
such that magnetic field stability is not a concern for the resonance
widths and separations discussed here. Losses due to photon
scattering can be kept minimal by choosing light, such as that of a CO$_2$
laser, that is far off resonance with any molecular
transition. Estimating the heating rates for $I=10^9\,$W/cm$^2$
in terms of the atomic photon
scattering rates~\cite{GrimmAdv00},  we find the largest
heating rate, that of the alkali atom, to be only of the order of
$1\,$nK/s for a wavelength of $10\,\mu$m. 
Wavelengths in the near infrared, e.g., 1064$\,$nm or 1550$\,$nm, 
could also be employed. For the telecom wavelength, we find a heating
rate of the order of $10\,\mu$K/s. This should be
sufficiently low to allow for adiabatic ramps whereas for 1064$\,$nm
with heating rates below $1\,$mK/s, the experiment needs to be
conducted within 1$\,$ms, better adapted to rf
magnetoassociation~\cite{JulienneRMP10}. The actually required
intensities and associated heating rates for these wavelengths
might, however, be lower due to 
the dynamic instead of the static polarizabilities coming into play.
This will be studied in detail elsewhere~\cite{TomzaFRlong}. 

Compared to electric field control of FRs for 
polar molecules~\cite{KremsPRL06,LiPRA07,MarcelisPRL08}, our proposal
corresponds to more favorable experimental conditions. For diatomics
consisting of an alkali atom and Sr or Yb, we find 
electric fields of several hundreds kV/cm to be 
required. This clearly exceeds current experimental
capabilities. Compared to the permanent electric dipole moment coupling to
a dc electric field, the interaction of Eq.~\eqref{eq:V} contains
diagonal in addition to off-diagonal matrix elements in the basis of
field-free rotational eigenstates. This explains the large shifts in
level positions which allow in particular for mechanism (i), i.e.,
$\delta\chi\to 0$. Moreover, the permanent dipole moment vanishes as
$1/r^7$ compared to the asymptotic $1/r^3$ behavior of the
polarizability. These facts together explain the much better prospects in
terms of experimental feasibility of our approach. 

In conclusion, we have shown that non-resonant light can be used to control
FRs of mixtures of open-shell/closed-shell atoms,
engineering their widths to reach up to a few Gauss. Such resonances
are sufficiently broad for magnetoassociation. The required field
strengths and control are all within current experimental
capabilities. Our proposal opens the way for producing ultracold
molecules with sizeable electric and magnetic dipole moments and thus
for many-body quantum simulations with such particles.

\begin{acknowledgments}
  We would like to thank Piotr \.{Z}{}uchowski and Axel G\"orlitz for making
their RbYb potentials available to us prior to publication and Michael
Drewsen and Jun Ye for providing their expertise on the state of the
art of laser technology.   
The authors enjoyed hospitality of the Kavli Institute of Theoretical
Physics where the work reported here was initiated. 
MT is supported by the Foundation for Polish Science MPD Programme
co-financed by the EU European Regional Development Fund.    RGF
acknowledges a Mildred Dresselhaus award from the excellence cluster
'The Hamburg Center for Ultrafast Imaging -- Structure, Dynamics and
Control of Matter at the Atomic Scale' of the
Deutsche Forschungsgemeinschaft.
Financial support by the Spanish project FIS2011-24540 (MICINN), the
Grants P11-FQM-7276 and FQM-4643 (Junta de Andaluc\'{i}a), and
Andalusian research group FQM-207, the project N-N204-215539 of 
the Polish Ministry of Science and Higher Education, 
the MISTRZ program of the Foundation for Polish Science,
and in part by the National Science Foundation under Grant No. NSF
PHY11-25915 is gratefully acknowledged.
\end{acknowledgments}


\end{document}